\documentstyle[aps,psfig,epsfig,amssymb,twocolumn]{revtex}
\begin{document} \draft
\title{Maximum-likelihood reconstruction of CP maps}
\author{Massimiliano F. Sacchi}
\address{Optics Section, Blackett Laboratory, 
Imperial College London, London SW7 2BZ, England \\
and Dipartimento di Fisica `A. Volta', Universit\`a di Pavia 
and Unit\`a INFM,  via
A. Bassi 6, I-27100 Pavia, Italy}
\maketitle
\begin{abstract}
We present a method for the determination of the completely 
positive (CP) map describing a physical device 
based on random preparation  of the input states, random measurements
at the output, and 
maximum-likelihood principle. In the numerical implementation 
the constraint of completely positivity 
can be imposed by exploiting the isomorphism between linear
transformations from Hilbert spaces $\cal H$ to $\cal K$ 
and linear operators in ${\cal K}\otimes {\cal H}$. The effectiveness of the
method is shown on the basis of some examples 
of reconstruction of CP maps related to 
quantum communication channels for qubits.
\end{abstract}

\pacs{PACS Numbers: 03.65.-w, 03.67.-a, 03.67.Hk}
The problem of characterizing 
a physical device in the quantum domain and of 
reconstructing input-output relations has been recently addressed 
in a number of papers \cite{macca,qcmc}. These issues are obviously
interesting for technological purposes. For example, 
the practical determination of 
the transformations acting on quantum states is of great 
relevance in the new fields  of quantum information, computation and
cryptography \cite{lo}. 
In this realm physical objects as optical fibers,  
parametric  amplifiers, directional couplers, C-Not gates, quantum
cloning machines, quantum 
communication channels, etc., should be characterized
with very high precision.   
From the theoretical point of view 
a physical device is described in terms of a completely positive
(CP) map. Hence, an experimental method of reconstructing a CP map
would lead to a complete characterization. On the other hand, 
an effective and reliable technique 
for the determination of a CP map could allow to check experimentally the 
correctness of the theoretical assumptions made in the description of 
the physical device. Finally, 
recall that the structure of CP maps naturally emerges in the theory 
of open systems \cite{open}. 
It follows that an experimental technique to estimate a
CP map also allows to investigate the interaction between different
systems, typically a system-reservoir interaction. 
\par Here in this letter we consider the general problem of
reconstructing the CP map related to  a physical device, without any
assumptions on its mathematical form. 
We propose a method which resembles ordinary tomography for the use of 
random quantum measurements at the output, 
but also employs random input states, 
in order to have the richest statistics. The
maximum-likelihood principle is then applied, using a suitable
parameterization of CP maps which  
is allowed by the isomorphism between linear
transformations from Hilbert spaces $\cal H$ to $\cal K$ 
and linear operators in ${\cal K}\otimes {\cal H}$ \cite{depi,jami}. 
The maximum-likelihood method 
has been used in the context of phase measurement \cite{phase}, and  
to estimate the density matrix \cite{matrix} and some parameters of
interest in quantum optics \cite{qcmc}.  
\par The maximum-likelihood principle says  that the best
estimation of some unknown 
parameters is given by  the values that are  most likely to generate 
the data one experimenter 
observes. This principle is quantified by a maximum search of a
functional of the unknown parameters that corresponds to (the 
logarithm of) the theoretical probability of getting the data 
one has collected. 
In the following we derive the likelihood functional ${\cal
L}( {\cal E})$ that  links the experimental outcomes with the unknown
CP map $\cal E$ that characterizes the device. 
We consider a sequence of $K$ independent measurements on the output, 
each described by a POVM $F_l(x_l)$, where $x_l$ denotes the
outcome at the $l$th measurement, and  $l=1,2,...,K$. We denote by
$\rho _l$ the state at the input at the $l$th run. 
The probability of getting the string of outcomes
$\vec x=\{x_1,x_2,...,x_K\}$ is then given by 
\begin{eqnarray}
p(\vec x)=
\Pi _{l=1}^K \hbox{Tr}[{\cal E} (\rho _l) F_l (x_l)]\;.\label{prod}
\end{eqnarray}
The maximum-likelihood principle states that the best estimate of the
map $\cal E $ maximizes the expression in Eq. (\ref{prod}) 
over the set of completely positive maps. 
More conveniently, one can search the maximum of the logarithm of
Eq. (\ref{prod}), namely
\begin{eqnarray}
{\cal L} ({\cal E})=\sum _{l=1}^K \log 
\hbox{Tr}[{\cal E} (\rho _l) F_l (x_l)]\;.\label{log}
\end{eqnarray}
The likelihood functional ${\cal
L}( {\cal E})$ is  concave, and in the present case it is
defined on a convex set---the set of CP maps. It follows that the
maximum is achieved by a single CP map, or by a closed convex subset
of CP maps. In the last case one can infer that the data sample is
not  sufficiently large, or the set of measurements is not a {\em
quorum}. 
\par The maximization problem is constrained by 
completely positivity and trace-preserving properties  of the
map $\cal E$. A trace-preserving CP 
map is a linear map from operators in Hilbert space ${\cal H}$ 
[dim(${\cal H})=N$] to 
operators in $\cal K$ [dim(${\cal K})=M$]
which can be written equivalently as
follows 
\begin{eqnarray}
{\cal E}(\rho)&=&\sum_k A_k\,\rho\,A_k^{\dag } \label{ss}\\ 
&=&\hbox{Tr}_{\cal H}[(\openone _{\cal K}\otimes \rho ^T) \,S] \label{trk}
\\
&=&\sum_{n=1}^{N^2}p_n\,U_n \,\rho \,
U_n^{\dag }\;,\label{us}
\end{eqnarray}
where 
\begin{eqnarray}
&&\sum_k A_k^{\dag }A_k = \openone _{\cal H}  \;,\\
&& S \geq 0 \quad \hbox{and }\quad \hbox{Tr}_{\cal K}[S]=\openone _{\cal
H} \;,
\label{sus}
\\
&&\hbox{Tr} [U_i^\dag \,U_j ]=\delta _{ij}\quad \hbox{and }\quad 
\sum_{n=1}^{N^2}p_n\, U_n^\dag U_n= \openone _{\cal H} 
\;,\label{sup}
\end{eqnarray}
and $T$ denotes the transposition. 
Eq. (\ref{ss}) is the well known Kraus decomposition. Eq. 
(\ref{trk}) exploits the isomorphism between linear
maps from ${\cal H}$ to ${\cal K} $ and linear
operators on the tensor-product space ${\cal K} \otimes {\cal H}$ 
\cite{depi,jami,xu,nota}.    
The operator $S$ can be written in terms of the map $\cal E$ as follows
\begin{eqnarray}
S=\sum_{i=1} ^{N^2} {\cal E}(V_i)\otimes V_i^* \;\label{s}
\end{eqnarray}
where $*$ denotes the complex conjugation, 
and $\{V_i\}$ is any orthonormal basis for the space of linear
operators on $\cal H$, namely
\begin{eqnarray}
\hbox{Tr}[V_i^\dag V_j]=\delta_{ij}
\;,\label{on1}
\end{eqnarray}
and for any operator $O$
\begin{eqnarray}
O=\sum_{i=1}^{N^2}\hbox{Tr}[V_i^\dag \,O]V_i\;.\label{on2}
\end{eqnarray}
Notice also that for linearity \cite{xu,lop}
\begin{eqnarray}
S ={\cal E} \otimes \openone (|\Psi \rangle
\langle \Psi |)\;,\label{smax}
\end{eqnarray}
where $|\Psi \rangle $ represents the 
(unnormalized) maximally entangled state $|\Psi \rangle =
\sum _{n=1}^N |n \rangle |n \rangle $.  
\par\noindent Eq. (\ref{trk}) can also be written as
\begin{eqnarray}
{\cal E} (\rho)= \hbox{Tr}_{\cal H}[(\openone _{\cal K}\otimes \rho ) 
S ^\Gamma] \label{trk2}
\;,
\end{eqnarray}
where $\Gamma $ denotes the partial 
transposition in ${\cal H}$, and then 
\begin{eqnarray}
S ^\Gamma  
=\sum_{i=1} ^{N^2} {\cal E}(V_i)\otimes V_i^\dag  \;.\label{sg}
\end{eqnarray}
\par\noindent Finally, Eq. (\ref{us}) can be shown as follows. 
Chosen an orthonormal basis $\{V_i\}$ in the sense of
Eqs. (\ref{on1}) and (\ref{on2}), 
Eq. (\ref{ss}) rewrites ${\cal E}(\rho)=\sum_{i,j=1}^{N^2}
q_{ij}V_i\rho V_j^\dag $, with $q_{ij}=\sum_k \hbox{Tr}[A_kV_i^\dag ]
\hbox{Tr}[A^\dag _kV_j]$. The matrix $Q$ with element $q_{ij}$ is
positive, and it can be diagonalized $Q=WDW^\dag $ with $W$
unitary. Then one has 
\begin{eqnarray}
{\cal E}(\rho)=\sum_{i,j=1}^{N^2} (WDW^\dag )_{ij}\,V_i\,\rho \,V_j^\dag =
\sum _{n=1}^{N^2} p_n\, U_n \,\rho \,U_n^\dag \;,\label{result}
\end{eqnarray}
where $p_n= (D)_{nn} \geq 0$ and $\{U_n\}$
 is the new orthonormal basis $U_n =\sum_{i=1}^{N^2} (W)_{ik}V_i$. 
\par For ${\cal H}\equiv{\cal K}$ the matrices $A_k$ and $U_n$ are
squared. When referring to quantum communication channels, 
the channel is called bistochastic if also 
$\sum_k{ A_k A_k^\dag }=\openone _{\cal H}$. Moreover, 
operators $\{U_n \}$ in Eq. (\ref{us}) could  be unitary, 
and in such case the (bistochastic) 
channel is said to be given by external random 
fields. For a qubit system (${\cal H }={\cal C}^2$) the set of 
bistochastic and external-random-field channels coincide \cite{otto}.  
\par The isomorphism between linear maps and operators established 
in Eqs. (\ref{trk})--(\ref{s}) or (\ref{trk2})--(\ref{sg})   
has been useful for the 
study of positive maps \cite{xu,nota}, and to address the problem of
separability of CP maps \cite{sepa}. 
For our task---the maximization of ${\cal L}({\cal E})$---it is also
crucial. The condition $S\geq 0$ allows one to write 
\begin{eqnarray}
S=C^\dag C\;,\label{tt}
\end{eqnarray}
where $C$ is an upper triangular matrix. Moreover, the 
diagonal elements of $C$ can be chosen as positive.  
 Such decomposition---referred to as Cholesky
decomposition---is commonly used in linear programming
\cite{ciarlet}. Similarly, one has for  
the matrices $\rho_l^T$ and the POVM's
$F_l(x_l)$ 
\begin{eqnarray}
\rho_l^T=R^\dag _l R_l \;,\qquad F_l(x_l)=A^\dag _l(x_l)A_l(x_l)\;.\label{deco}
\end{eqnarray}
From Eqs. (\ref{trk}), (\ref{tt})  and  (\ref{deco}), 
the log-likelihood functional in Eq. (\ref{log}) rewrites
\begin{eqnarray}
{\cal L}({\cal E})&\equiv &{\cal L}(C)=
 \sum _{l=1}^K \log 
\hbox{Tr}[C^\dag C  (R^\dag _l R_l \otimes 
A^\dag _l(x_l)A_l(x_l))] \nonumber \\& =&
 \sum _{l=1}^K \log \sum_{n,m=1}^{NM}\left| 
\langle \!\langle n |C (R^\dag _l \otimes A^\dag _l (x_l)) |
m \rangle \!\rangle  
\right|^2\;,\label{log2}
\end{eqnarray}
where $\{|n \rangle \!\rangle \}$ denotes an orthonormal basis 
for ${\cal H}\otimes {\cal K}$. The expression obtained in
Eq. (\ref{log2}) for the likelihood functional automatically 
satisfies the constraint of completely positivity. Furthermore, 
the terms appearing as argument of the logarithm are explicitly
positive, thus assuring the stability of 
numerical methods to evaluate ${\cal L}(C)$. 
\par The trace-preserving condition is given in terms of the matrix $S$ as
$\hbox{Tr}_{\cal K}[S]=\openone _{\cal H}$. 
This can  be taken into account by using the method of 
undetermined Lagrange multipliers, then maximizing 
\begin{eqnarray}
{\cal L}'({\cal E})&=&
{\cal L}({\cal E})
-\sum_{i,j=1}^N \mu _{ij} \langle i|\hbox{Tr}_{\cal K}[S]|j
\rangle  \nonumber \\&= & 
{\cal L}({\cal E})-\hbox{Tr}[(\openone _{\cal
K}\otimes \mu )\, S]\;,
\end{eqnarray}
where $\mu $ is the undetermined matrix $\mu =\sum_{i,j=1}^N
 \mu _{ij } |i \rangle \langle j|$. The multipliers $\mu _{ij}$ cannot be
easily inferred, except  the condition $\hbox{Tr}[\mu] =K$. Writing 
$S$ in terms of its eigenvectors as 
$S=\sum_{i} s_i^2 |s_i \rangle \!\rangle  \langle \!\langle s_i|$ , 
the maximum likelihood condition $\partial {{\cal L}'}({\cal 
E})/\partial s_i=0$ implies  
\begin{eqnarray}
&&\sum _{l=1}^K \frac{\hbox{Tr}[(\rho _l^T\otimes F_l (x_l))
s_i\, |s_i \rangle \!\rangle  \langle \!\langle s_i|]}
{\hbox{Tr}[(\rho_l^T\otimes F_l (x_l)) S ]}
\nonumber \\&&=
\hbox{Tr}[(\openone _{\cal K}\otimes \mu )
s_i\, |s_i \rangle \!\rangle  \langle \!\langle s_i|]
\;.\label{mu}\end{eqnarray}
Multiplying by $s_i$ and summing over $i$ gives $\hbox{Tr}[\mu ]=
K$. However, notice that the constraint $\hbox{Tr}[S]=N$ which follows from 
$\hbox{Tr}_{\cal K}[S]=\openone _{\cal H}$ 
isolates a closed convex subset of the set of positive 
matrices. Hence, the maximum of the concave likelihood functional still 
remains unique under this looser constraint,  and one can check {\em 
a posteriori} that 
the condition $\hbox{Tr}_{\cal K}[S]=\openone_{\cal H}$ is fulfilled. 
The functional we maximize is then 
\begin{eqnarray}
{\tilde {\cal L}}(C)={\cal L}(C) - 
\frac K N \hbox{Tr}[C^\dag C]\;,\label{tilde}
\end{eqnarray}
where ${\cal L}(C)$ is given in Eq. (\ref{log2}), and the value of the 
multiplier has been obtained through a derivation similar to 
Eq. (\ref{mu}).  
The number of unknown real parameters is given by $(NM)^2$. The
problem of 
maximization of functionals as Eq. (\ref{tilde}) enters the realm of
programming and numerical algebra optimisation, where various
techniques 
are known \cite{ciarlet}.  
\par In the following we show the effectiveness of our method on the
basis of some examples of quantum communication channels for qubits,
i.e. ${\cal H}={\cal K}={\cal C}^2$. In this case $(NM)^2=16$, and for such 
a relatively small number of parameters one can efficiently apply 
the method of downhill simplex \cite{ciarlet,simplex} to find the maximum of 
the log-likelihood functional.  
This method has been reliable in the reconstruction of the 
density matrix of radiation field and spin systems \cite{matrix}. 
The results in the following simulations have been obtained using 
random pure  states at the input of the channel, along with a projective 
measurement in a random direction at the output. 
\par The first example is the Pauli channel for qubits
\begin{eqnarray}
{\cal E}_p(\rho )= \sum_{i=1}^4 p_i\,\sigma _i \rho \sigma _i\;,
\label{pauli}
\end{eqnarray}
where $\sum_{i=1}^4 p_i=1$, $\sigma _0\equiv \openone $, and $\sigma
_i$ ($i\!=\!1,2,3$) denote  the customary Pauli matrices. From
Eq. (\ref{smax}) the corresponding positive matrix $S_p$ writes 
\begin{eqnarray}
S_p=  \left( 
\begin{array}{cccc}
 p_0+p_3 & 0 & 0 &p_0-p_3\\
 0 & p_1+p_2 &p_1-p_2  &0\\
 0& p_1-p_2 &p_1+p_2&0 \\
p_0-p_3 & 0 &0&p_0+p_3
\end{array} \right )
\;,\label{sp}
\end{eqnarray}
on the lexicographically ordered basis 
$|00 \rangle$, $|01 \rangle$, $|10 \rangle$, $|11 \rangle $, where $|0
\rangle $ and $|1 \rangle  $ corresponds to the eigenstates of $\sigma _z$
with eigenvalues $1$ and $-1$, respectively.  
\par In table \ref{t:uno} we reported the reconstructed matrix
elements of $S_p$ as obtained by a Monte Carlo 
simulation with $K=30000$ runs, 
for theoretical values $p_0=0.3$, 
$p_1=0.2$, $p_2=0.4$, and $p_3=0.1$. 
The trace-preserving property corresponds 
to the conditions $S_p(1,1)+S_p(3,3)=1$, $S_p(2,2)+S_p(4,4)=1$, and 
$S_p(1,2)+S_p(3,4)=0$, which are clearly satisfied. 
The estimated values compare very
well with the theoretical ones.
\par
\begin{figure}[hbt]
\vskip .5truecm
\begin{center}
\epsfxsize=.5 \hsize\leavevmode\epsffile{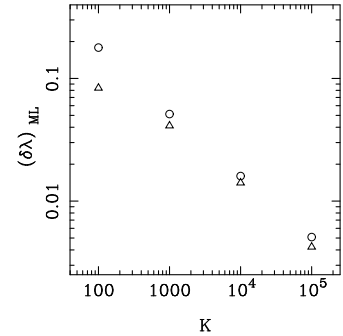}
\end{center}
\caption{Maximum-likelihood estimation of the CP map related to the
depolarizing channel. The picture shows the value of the statistical
error $(\delta \lambda )_{ML}$ 
in the estimation of the parameter $\lambda $ (theoretical value 
$\lambda  =0.8$) versus the size $K$ of the data sample. Circles referred
to a ML reconstruction without assumptions on the form of the
CP map; triangles are the results when assuming the
external-random-field form. 
The asymptotic  dependence of the statistical error versus $K$ 
is inverse square-root $(\delta \lambda
)_{ML}\propto K^{-1/2}$, as it is demanded by the central limit theorem.}
\label{f;depo}\end{figure}

 For $p_1=p_2=p_3$ in Eq. (\ref{pauli}), one obtains the 
depolarizing channel
\begin{eqnarray}
{\cal E}_d(\rho )= \lambda \rho + \frac {1-\lambda }{2} \openone
\;,\label{depo}
\end{eqnarray}
with $\lambda =1-4p_1$. In Fig. \ref{f;depo} (circles) we reported the
statistical error $(\delta \lambda )_{ML}$ 
in the evaluation of the parameter $\lambda $ versus 
the size $K$ of data sample (with theoretical value 
$\lambda =0.8$). The value of $\lambda _{ML}$ has been inferred by the 
complete reconstruction of the matrix $S_p$. However, notice that one 
can also implement the maximum-likelihood method upon assuming the
form of the CP map 
as in Eqs. (\ref{pauli}) or (\ref{depo}). In such case the
space of parameters is reduced to 4 and 1,
respectively. Fig. \ref{f;depo} also shows the results obtained by a
4-parameters estimation (triangles), thus by assuming an
external-random-field channel. 
In both cases one has an  asymptotic inverse square-root dependence of the
statistical error on the size of the data sample $(\delta \lambda
)_{ML}\propto K^{-1/2}$, according to 
the central limit theorem. 
\par
\begin{figure}[hbt]
\vskip .5truecm
\begin{center}
\epsfxsize=.5 \hsize\leavevmode\epsffile{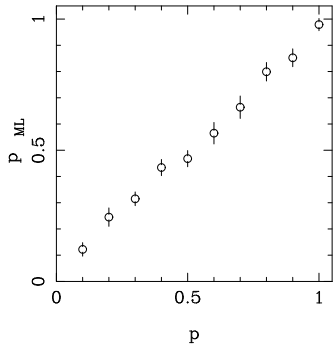}
\end{center}
\caption{Maximum-likelihood reconstruction of the CP map describing an 
amplitude damping channel. The value of the 
parameter $p$ is inferred by reconstructing 
the positive matrix $S_a$ in Eq. (\ref{sa}). 
Random pure states at the input and projection
along random directions at the output have been used, with  
$K=10000$ measurements.}
\label{f;damp}\end{figure}

 In the last example we consider a non-bistochastic channel, namely 
the amplitude damping channel 
\begin{eqnarray}
{\cal E}_a(\rho )=M_1\,\rho \,M_1 + M_2\,\rho \,M_2  
\;\label{d1}
\end{eqnarray}
with 
\begin{eqnarray}
M_1=\left(\begin{array}{cc}
 1 & 0  \\
 0 & \sqrt{p}
\end{array} \right ) \quad \hbox{and }\quad 
M_2=\left(\begin{array}{cc}
 0 & \sqrt{1-p}  \\
 0 & 0
\end{array} \right ) 
\;.\label{d2}
\end{eqnarray}
The corresponding positive matrix $S_a$ write 
\begin{eqnarray}
S_a =\left( 
\begin{array}{cccc}
 1&0&0&\sqrt p\\
 0 & 1-p&0 &0\\
 0&0 &0&0 \\
\sqrt p &0&0&p
\end{array} \right )
\;.\label{sa}
\end{eqnarray}
In Fig. \ref{f;damp} we have plotted the estimated value $p_{ML}$ 
of parameter $p$ its theoretical value, as inferred by the
reconstruction of the matrix $S_a$ through 
$K=10000$ random measurements.
\par In conclusion, we have proposed a method for reconstructing 
the completely positive map related to a physical device, 
based on the maximum likelihood principle. The
method is very general, does not require {\em a priori} knowledge of
the mathematical structure of the CP map, 
and can be adopted in many fields as 
quantum optics, spin systems, optical lattices, 
atoms, etc. We have shown some examples of reconstruction 
of CP maps related to quantum communication channels, applying the
downhill simplex method for the search of the maximum of the
likelihood functional. 

\par\noindent {\em Acknowledgments.} The author 
would like to thank the Leverhulme Trust foundation for partial
support. This work has been supported by European Program EQUIP and by 
the Italian Ministero 
dell'Universit\`a e della Ricerca Scientifica e Tecnologica (MURST) 
under the co-sponsored project 1999 {\em Quantum Information
Transmission and Processing: Quantum Teleportation and Error Correction}. 

\begin{table}[h]\begin{center}
\begin{tabular}{|r|r|r|}
$i$ & $j$  & $S_p(i,j)$  \\ \hline
1& 1 & (0.388964638,0.) \\ \hline
 1& 2 & (-0.011561621,-0.0160863415)         \\ \hline       
1& 3& (-0.00103390675,-0.0164688228)\\ \hline
1& 4& (0.188891975,-0.0241343938)\\ \hline
 2& 2 &(0.617439461,0.)\\ \hline
 2& 3 &(-0.182118262,0.000703314322)\\ \hline
 2 &4 &(-0.00825923682,0.020653044)\\ \hline
 3 &3 &(0.606198593,0.)\\ \hline
 3 &4 &(0.00111897098,0.0150693168)\\ \hline
 4 &4 &(0.389230293,0.)
\end{tabular}
\end{center}
\caption{Maximum-likelihood estimation of real and imaginary
parts of the matrix elements of $S_p$ related to 
the Pauli channel in Eq. (\ref{pauli}),   
for theoretical values $p_0=0.3$, 
$p_1=0.2$, $p_2=0.4$, and $p_3=0.1$. Random pure states and projective 
measurements along random directions have been used, with $K=30000$
runs. Compare with the theoretical values as obtained from Eq. (\ref{sp}). 
The statistical error in the estimation of the matrix elements is 
around $0.01$. For typical values and behavior of the statistical 
errors, see Fig. \ref{f;depo}.\label{t:uno}}
\end{table}

\begin{thebibliography}{99}
\bibitem{nota} For a positive---but not completely positive---map 
the condition of positivity of the matrix $S$ is relaxed, by only requiring 
positivity for  tensor product of vectors, namely  
$\langle \phi | \langle \psi |S |\phi \rangle |\psi \rangle  \geq 0$
(see. Refs. \cite{jami,xu})
\end{thebibliography}
\end{document}